# Advancing Brain-Machine Interfaces: High Data Rate Battery-Free Implants

Aminolah Hasanvand
Department of electronic systems (IES)
*Norwegian University of Science and Technology*
Trondheim, Norway
aminolah.hasanvand@ntnu.no

Ali Khaleghi
Department of electronic systems (IES)
*Norwegian University of Science and Technology*
Trondheim, Norway
ali.khaleghi@ntnu.no

Cyril Beguet
Blackrock Microsystems Europe GmbH
Hannover, Germany
cbeguet@blackrockneuro.com

Paul Wanda
Blackrock Microsystems Europe GmbH
Hannover, Germany
pwanda@blackrockneuro.com

Ilangko Balasingham
Department of electronic systems (IES)
*Norwegian University of Science and Technology*
Trondheim, Norway
ilangko.balasingham@ntnu.no

***Abstract*— Implantable wireless brain-machine interfaces (BMI) encounter significant challenges in miniaturization, power consumption, and high data volume. While systems utilizing high resolution microelectrode arrays offer precision brain readout and/or stimulation, achieving high-rate wireless connectivity (32-128 Mbps) consumes excessive power, unsuitable for long-term use with implant batteries. This paper addresses wireless connectivity and power challenges by employing radio frequency backscatter and near-field wireless charging. This approach eliminates transceiver electronics in the implantable, reducing implant power consumption by offloading complexity to off-body reader electronics. It enables wireless powering of implantable neural recording and stimulation chips through magnetic coupling, enabling a fully implantable brain-machine interface. We present preliminary test results for this design scenario, demonstrating the feasibility of our approach.**

***Keywords*— *Implantable wireless brain-machine interfaces, multielectrode arrays, near-field wireless charging, wireless radio frequency backscatter***

## I. Introduction

BMI (Brain-Machine Interface) describes a technology that enables communication between the brain and an external device, such as a computer or a machine. High-resolution invasive BMI interfaces have been used to create a direct connectivity pathway between the brain and prosthetic devices for individuals with disabilities [1]. The state-of-the-art intracortical BMIs involve implanting electrode arrays directly into the brain cortex to record and decode neural signals or stimulate the neurons [2]. Microelectrode arrays, such as the Blackrock Utah Electrode Array[3] among others, often use small electrodes that penetrate the brain tissue and are in close connectivity with the neurons or microcontacts distributed on the cortical surface, to generate high spatio-temporal resolution allows for precise and detailed neural signal detection and feedback that cannot be attained with wearable EEGs [4]. However, when recording the spiking activity of individual neurons, the needed combination of high sampling rates and high data resolution can generate data up to 1 Mbps per channel. Transfer of this data is currently realized with a direct wire connection to a pedestal on the skull [1], but this transcutaneous approach poses challenges for device longevity, care, and safety. Invasive, implanted BMI devices are likely more stable and reliable over time and would be suggested for long-term use [2]. Such implantable devices need to be small in size, ultra-low power, safe, and secure, Here we focus on the data readout and communication to such implants. Transmitting these data over a wireless link can eliminate the pedestal that is prone to infection with physical implications that reduce patient mobility and social interactions.

Providing a near-zero power wireless connection for BMI implants is the goal of this project. In past efforts, the initial power consumption estimation for a BMI using a microelectrode array of 100 electrodes with 1 kHz sample per channel is 50-100 mW(milli watts) for the local neural recording chip without a communication module [5]. However, available technologies can reduce the power requirement to the order of 6.5 mW [6]-[7]. Wireless backscatter is an approach that can eliminate the communication electronics from the implant and bring the communication power to zero for the implant unit by shifting complexity to the off-body wearable and reduce the electronics thermal side effects. In [8]-[9], we previously demonstrated the use of wireless backscatter communication (10 Mbps) for a video capsule endoscopy system without using a transceiver in the capsule. We extend the concept to the requirements of a modern BMI implant using the wireless RF backscatter approach for data readout with simultaneously wirelessly charging, further enabled by the removal of power-hungry transceiver components from the implant. We demonstrated the feasibility of these technologies in an implantable form factor with 24 Mbps data rate and simultaneous charging capacity. The overall concept of this work is depicted on Fig. 1.

## II. Backscatter communication

Here, we consider the design of a wireless platform to support the transmission of neural spiking activity using backscatter load modulation. The implanted neural interface is assumed as a device consisting of a microelectrode array and electronics including an amplification/digitization neural recording chip. The digital data are emulated by a micro-controller and an FPGA. The digital data stream is read with backscatter load modulation. In this model, the data produced for each channel of the microelectrode array is estimated at 1 Mbps, which for a typical 32-channel array becomes 32 Mbps. Although data compression can reduce the channel bandwidth, it would also add power consumption, by requiring the involvement of a processing unit with a high clock rate. Therefore, we do not apply data compression in this work. In the prototype implementation we use BPSK load modulation while with QPSK load modulation we can reduce the data bandwidth without affecting the signal to noise ratio (SNR). Thus, a data rate of up to 64 Mbps can be supported. The digitized neural data stream is sent in serial format to an RF switch that controls the radar cross section of the implant antenna, where the switching alters the differential RCS of the antenna. The switch operates between the short and open circuit states to provide two different RCS and phase values in the radar link. An external reader system, with exciter antenna and receiver antenna connected to the demodulation device decodes the backscatter data with a few microseconds latency.

## III. Antennas and SAR Limitation

The design of compact antennas for wireless backscatter presents several challenges. The primary constraint is the mutual coupling between the antennas in a bi-static radar scenario, as well as achieving antenna matching in the transmitter and receiver. The implant antenna needs to be compact, planar, and easily integrated into a BMI casing. Furthermore, the design must consider important criteria such as external antenna SAR (Specific Absorption Rate) and interference to BMI electronics. Given the implant antenna's close proximity to tissues, the effects of biological tissues must be taken into account during antenna design. Thermal effects, including those resulting from the external transmitter antenna and reflections from the implant and casing, must be kept below 2 W/kg. Additionally, the implant antenna should maximize radar cross-section and be controlled by the implant switching mechanism to enable the transmission of digital information to an external reader. The external reader transmits a continuous wave signal during operation and records reflections from the implant. Compliance with RF radiation standards is essential for emission approvals. Although the backscatter data is wideband and the received signal is susceptible to RF interference, the emission of the wideband data signal is confined to head proximity, making it unlikely to be decoded by other receivers. In this design example, RF emission by an external transmitter at 434 MHz with an emission bandwidth of below 1 kHz is used. The transmitter and receiver antennas operate in the implant near-region, so the coupling among the antennas is near-field. The implant antenna uses a dual-polarized scheme to compensate for the polarization mismatch. The external transceivers are orthogonal circular polarized to limit the mutual coupling, resulting in 40 dB isolation for the antennas on a circular surface of 40 mm. The implant antenna is developed in a dual polarized patch of size 20 mm diameter. The RF emission is 100 mW, and the distance between the external and subcutaneous implant on the subject's head is 5 mm. The full wave FDTD method is used to simulate the antennas, considering the realistic model of the human head using the HUGO model (Fig. 1). The computed RF SAR for the accepted of 100 mW power is 0.166 W/Kg. The BMI is powered by an induction coil operating at 13.56 MHz using the NFC (near field communication) protocol, delivering 30 mW of power to the implant. Due to the non-magnetic nature of biological tissues, the magnetic coil results in lower SAR. However, it does generate some heating on the external transmitter, which can be effectively dissipated in air. The implant is designed to be positioned subcutaneously beneath the skin, and the external antenna system is integrated to facilitate both the powering of the implant and the wireless backscatter communication.

## IV. Wireless Power and Two Way Communication

To transmit power wirelessly, as well as for two-way communication with the brain implant for telemetry and telecommand, a low power NFC protocol is implemented, though this might not be allocated for medical application however the availability of the electronics led us to validate the project concept. Fig. 2 shows the application of NFC for power transmission and for establishing a two-way link. By using this method and our antenna system setup we prove power delivery up to 30 milliwatts at the same time as sending and receiving telemetry and telecommand data up to 400 kbps.

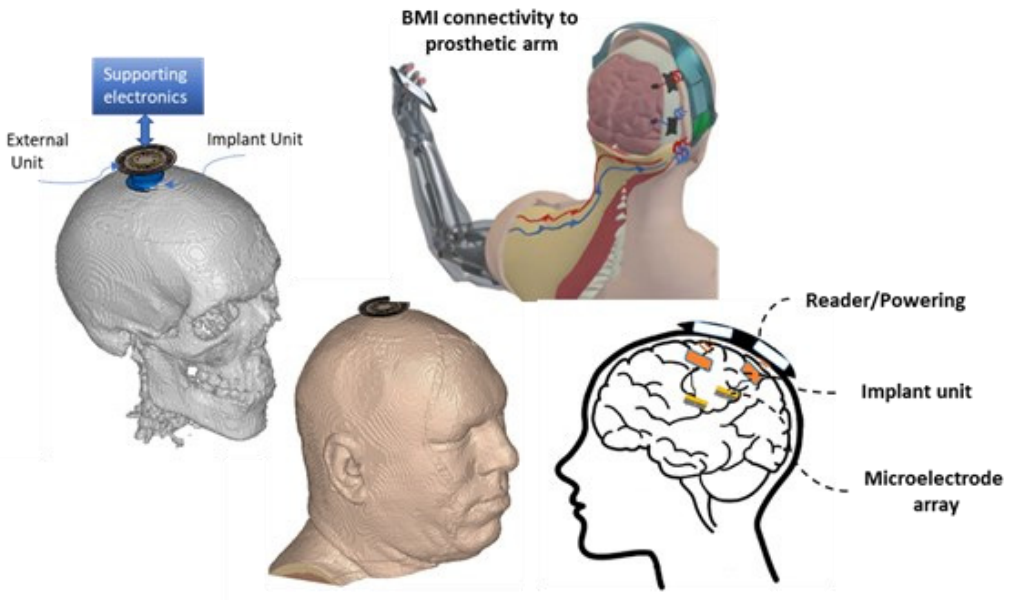

Fig. 1. Illustration of the simulation setup and placement of the implant and external reader and powering units.

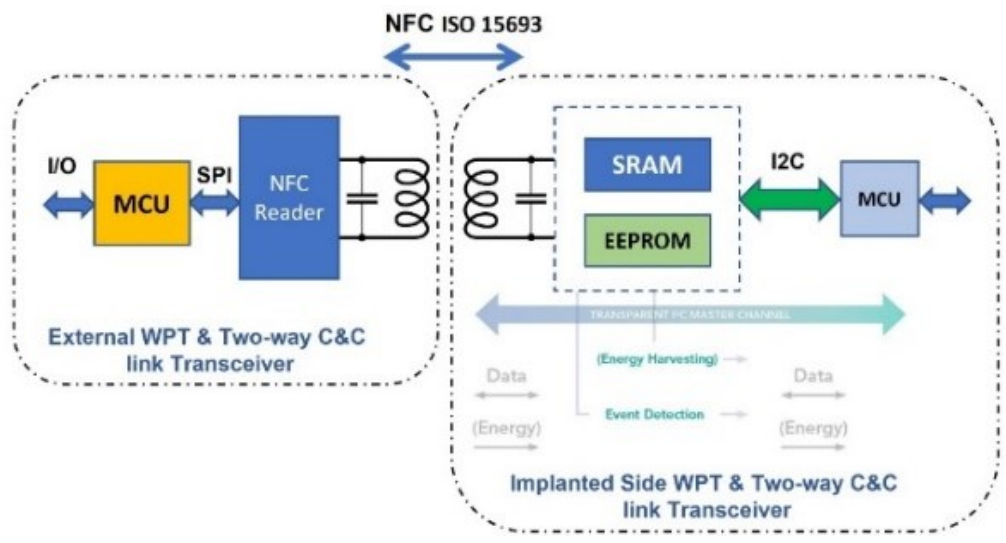

Fig. 2. WPT and Two-way command and control (C&C) communication link functional block diagram

## V. System Integration and Results

The envisioned design for the BMI prototype device incorporates readily available off-the-shelf components and an RF-transparent hermetic enclosure. The system's wireless components have been tested to validate its operation. The implant wireless hardware consists of three separate PCBs: a backscatter antenna, a wireless charging antenna, and the electronics. These PCBs are circular in shape, with a diameter of 25 mm and a thickness of 5 mm and are stackable (see 0). Ultra-low power electronics, such as the Max32 microcontroller and the ADG901 RF switch, are chosen, and BPSK load modulation is implemented. For testing, the microcontroller generates dummy data to control the switch states and ensure the system's functionality. Neural data, emulated by prerecorded waveforms, is streamed at a rate of 24 Mbps (while 32 Mbps is our ultimate goal) and used to control the switching process. In the final design, a microelectrode array and recording chip will be integrated in place of the microcontroller dummy data stream.

For command-control and wireless power transfer (WPT), a standard NFC RFID (Radio Frequency Identification) chip NTAG 5 is utilized. An NFC reader board, specifically the CLEV6630B, serves as the external component for WPT and two-way command/control (C&C) transceiver. This reader is designed to relay external commands, such as brain stimulation control, from an external device or sensors to the implant, while also receiving monitoring data from the implant through the NFC link. The external command data is transferred via NFC link, received by the implant microcontroller and validated. As a two-way wireless communication link, the NFC connection also facilitates the periodic transfer of housekeeping and condition monitoring data from the implant to the external transceiver using NFC's 13.56 MHz link. Furthermore, a received wireless power of 35 mW has been measured to be available for powering on-board electronics, such as a recording and stimulating chip. Power regulations have been implemented, including a 3-volt DC supercapacitor to power the chip.

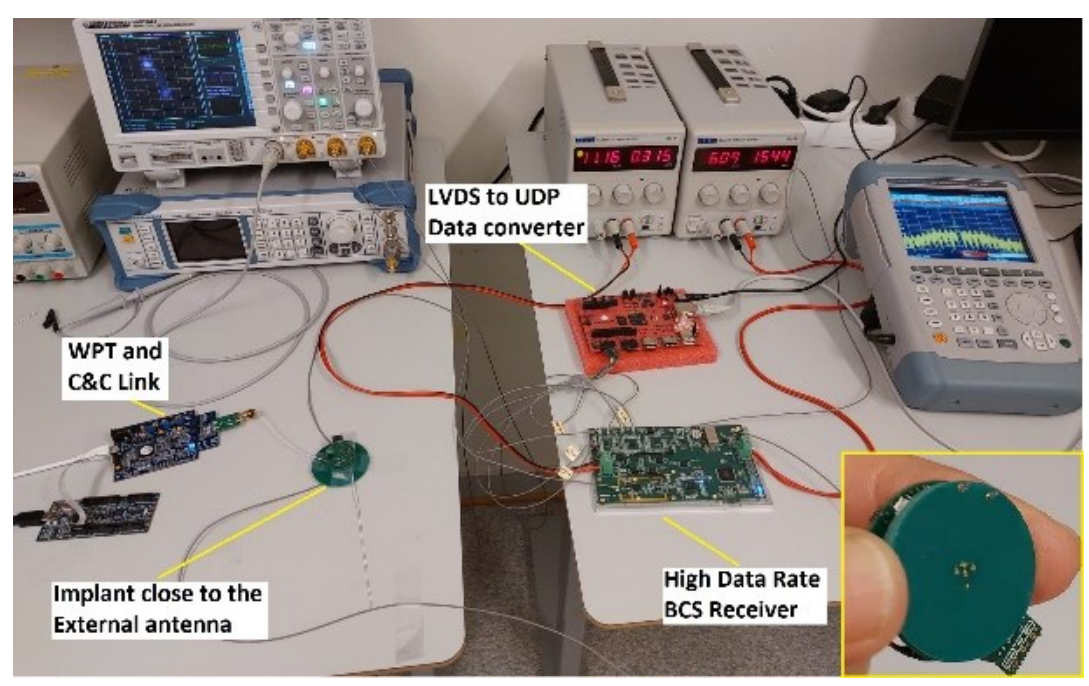

Fig. 3. Developed model of implanted part, integrated manufactured model. Integrated system test setup.

Fig. 3 shows the integrated system test of the aforementioned system in a loop. Dummy brain data is generated from an emulator at a rate of 24 Mbps and given to the implant board. These data are sent through the backscatter link. The implant part is placed at different distances from the external integrated antenna and according to the distance, the amount of generated power, and the performance of the implant is tested. The external backscatter modem decodes the received data of rate 24 Mbps, the clock and data recovery are performed and the data stream in LVDS (Low voltage differential signaling) fed to a UDP (User Datagram Protocol) network protocol and transferred to a PC for further processing. A software developed on Python to demodulate data in real-time and decode data of each channel. A video demonstration of the transmitted dummy data is generated to visually inspect the error rate of the system. To calculate the BCS link error and also evaluate the quality of the BCS link, first, in emulator 32 channels are generated, and for each of the thirty-two channels, a constant waveform is generated in such way that by receiving and detecting the waveforms and comparison with the sent data we calculate the quantitative error of the BCS link, while the data is shown in real time in a video. Fig. 4 shows received data, as a frame of video stream for each channel, from BCS link after being sent from the implant. These data are displayed in real time on the receiver computer. For the case where the implant is placed at a distance of 3 cm from the external antenna (in the air), and the external transmitter accepts 100 mW the data received from the BCS link has an error of $3.2\times10^{-3}$ (average BER).

## VI. Conclusion

A wireless communications and powering platform for a fully implantable BMI has been developed and demonstrated for wireless connectivity of up to 24 Mbps and simultaneous wireless powering up to 35 mW. The feasibility study was performed via simulations and benchtop demonstrations by wireless powering, streaming sample brain data and two-way telemetry and telecommand communication.

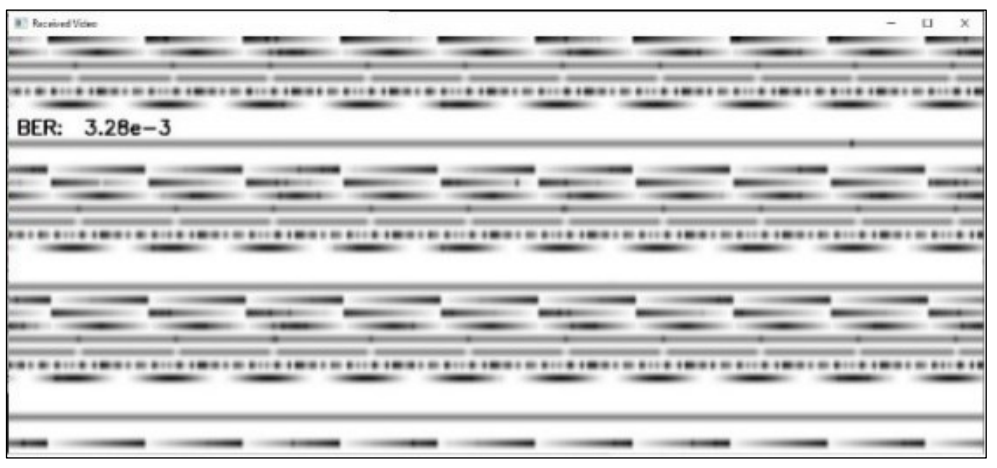

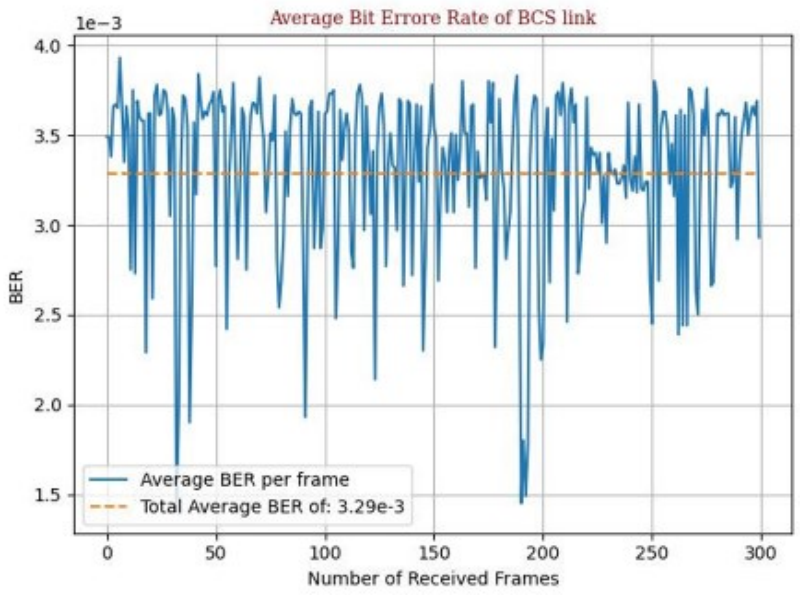

Fig. 4. Up: Received backscattered data for 3 cm separation between implant and external integrated antenna. Each video frame includes 10 period of each channel waveform that any period includs 30 sample of 16 bit, represented as a row in image (32×300×16 bit=153600 bit). In this frame BER=$3.2\times10^{-3}$. Down: calculated BER of each received frame of 153600 bit.

## Acknowledgment

The work has been supported by the project Brain-Connected inteRfAce TO machineS (B-CRATOS), (https://www.b-cratos.eu) under grant 965044, Horizon 2020 FET-OPEN. We acknowledge collaborators at Blackrock Microsystems Europe GmbH.